\documentclass[preprintnumbers,amsmath,amssymb]{revtex4}

\usepackage{graphicx}
\usepackage{dcolumn}

\usepackage{bm}

\begin{document}

\title{The dark side of gravity: Modified theories of gravity}

\author{Francisco S. N. Lobo}
\email{flobo@cosmo.fis.fc.ul.pt}
\affiliation{Centro de Astronomia
e Astrof\'{\i}sica da Universidade de Lisboa, Campo Grande, Ed. C8
1749-016 Lisboa, Portugal}
\affiliation{Institute of Gravitation \&
Cosmology, University of Portsmouth, Portsmouth PO1 2EG, UK}

\date{\today}

\begin{abstract}

Modern astrophysical and cosmological models are faced with two
severe theoretical difficulties, that can be summarized as the
dark energy and the dark matter problems. Relative to the former,
it has been stated that cosmology has entered a `golden age', in
which high-precision observational data have confirmed with
startling evidence that the Universe is undergoing a phase of
accelerated expansion. Several candidates, responsible for this
expansion, have been proposed in the literature, in particular,
dark energy models and modified gravity, amongst others. One is
liable to ask: What is the so-called `dark energy' that is driving
the acceleration of the universe? Is it a vacuum energy or a
dynamical field (``quintessence'')? Or is the acceleration due to
infra-red modifications of Einstein's theory of General
Relativity?
In the context of dark matter, two observations, namely, the
behavior of the galactic rotation curves and the mass discrepancy
in galactic clusters, suggest the existence of a (non or weakly
interacting) form of dark matter at galactic and extra-galactic
scales. It has also been proposed that modified gravity can
explain the galactic dynamics without the need of introducing dark
matter.
We briefly review some of the modified theories of gravity that
address these two intriguing and exciting problems facing modern
physics.

\end{abstract}


\maketitle

\section{Introduction}

Cosmology is said to be thriving in a golden age, where a central
theme is the perplexing fact that the Universe is undergoing an
accelerating expansion~\cite{expansion}. The latter, one of the
most important and challenging current problems in cosmology,
represents a new imbalance in the governing gravitational
equations. Historically, physics has addressed such imbalances by
either identifying sources that were previously unaccounted for,
or by altering the governing equations. The cause of this
acceleration still remains an open and tantalizing question.

The standard model of cosmology has favored the first route to
addressing the imbalance, namely, a missing stress-energy
component. In particular, the dark energy models are fundamental
candidates responsible for the cosmic expansion (see Refs.
\cite{Copeland} for a review and references therein). A simple way
to parameterize the dark energy is by an equation of state of the
form $\omega\equiv p/\rho$, where $p$ is the spatially homogeneous
pressure and $\rho$ the energy density of the dark energy. A value
of $\omega<-1/3$ is required for cosmic expansion, as dictated by
the Friedmann equation $\ddot{a}/a=-4\pi G(p+\rho/3)$, and
$\omega=-1$ corresponds to a cosmological constant. A possibility
that has been widely explored, is that of quintessence, a cosmic
scalar field $\phi$ that has not yet reached the minimum of its
potential $V(\phi)$~\cite{Douspis}. A common example is the energy
of a slowly evolving scalar field with positive potential energy,
similar to the inflaton field used to describe the inflationary
phase of the Universe. In quintessence models the parameter range
is $-1<\omega<-1/3$, and the dark energy decreases with a scale
factor $a(t)$ as $\rho_Q \propto a^{-3(1+\omega)}$~\cite{Turner}.
A specific exotic form of dark energy denoted phantom energy, with
$\omega<-1$, has also been proposed \cite{Caldwell}, and possesses
peculiar properties, such as the violation of the energy
conditions and an infinitely increasing energy density.
However, recent fits to supernovae, cosmic microwave background
radiation (CMBR) and weak gravitational lensing data indicate that
an evolving equation of state crossing the phantom divide, is
mildly favored, and several models have been proposed in the
literature~\cite{Vikman,phantom-divide}. In particular, models
considering a redshift dependent equation of state, possibly
provide better fits to the most recent and reliable SN Ia
supernovae Gold dataset.

It is also interesting to test specific models that are motivated
by particle physics against the SN data, rather than trying to fit
the phenomenological equations of state. In a cosmological
setting, it has also been shown that the transition into the
phantom regime, for a single field is probably physically
implausible \cite{Vikman}, so that a mixture of various
interacting non-ideal fluids is necessary. If confirmed in the
future, this behavior has important implications for theoretical
models of dark energy. For instance, this implies that dark energy
is dynamical and excludes the cosmological constant and the models
with a constant parameter as possible candidates for dark energy.
An alternative model to dark energy is that of the generalized
Chaplygin gas (GCG), based on a negative pressure fluid, which is
inversely proportional to the energy density \cite{GCGmodel},
i.e., $p_{\rm ch}=-A/\rho_{\rm ch}^{\alpha}$, where $A$ and
$\alpha$ are positive constants. An attractive feature of this
model, is that at early times, the energy density behaves as
matter, $\rho_{\rm ch}\sim a^{-3}$, where $a$ is the scale factor,
and as a cosmological constant at a later stage, $\rho_{\rm
ch}={\rm const}$. This dual behavior is responsible for the
interpretation that the GCG model is a candidate of a unified
model of dark matter and dark energy \cite{Bilic}, and probably
contains some of the key ingredients in the dynamics of the
Universe for early and late times. All of these models present an
extremely fascinating aspect for future experiments focussing on
supernovae, CMBR and weak gravitational lensing and for future
theoretical research.

One may also explore the alternative viewpoint, namely, through a
modified gravity approach. A very promising way to explain these
major problems is to assume that at large scales Einstein's theory
of General Relativity breaks down, and a more general action
describes the gravitational field. The Einstein field equation of
General Relativity was first derived from an action principle by
Hilbert, by adopting a linear function of the scalar curvature,
$R$, in the gravitational Lagrangian density. However, there are
no a priori reasons to restrict the gravitational Lagrangian to
this form, and indeed several generalizations of the
Einstein-Hilbert Lagrangian have been proposed, including
``quadratic Lagrangians'', involving second order curvature
invariants such as $R^{2}$, $R_{\mu \nu }R^{\mu \nu }$, $R_{\alpha
\beta \mu \nu }R^{\alpha \beta \mu \nu }$, $\varepsilon ^{\alpha
\beta \mu \nu }R_{\alpha \beta \gamma \delta }R_{\mu \nu }^{\gamma
\delta }$, $C_{\alpha \beta \mu \nu }C^{_{\alpha \beta \mu \nu
}}$, etc \cite{early}. The physical motivations for these
modifications of gravity were related to the possibility of a more
realistic representation of the gravitational fields near
curvature singularities and to create some first order
approximation for the quantum theory of gravitational fields.
In this context, a more general modification of the
Einstein-Hilbert gravitational Lagrangian density involving an
arbitrary function of the scalar invariant, $f(R)$, was considered
in \cite{Bu70}, and further developed in \cite{Ke81,Du83,BaOt83}.
Recently, a renaissance of $f(R)$ modified theories of gravity has
been verified in an attempt to explain the late-time accelerated
expansion of the Universe. Earlier interest in $f(R)$ theories was
motivated by inflationary scenarios as for instance, in the
Starobinsky model, where $f(R)=R-\Lambda + \alpha R^2$ was
considered \cite{Starobinsky:1980te}.
In particular, it was shown that cosmic acceleration can be indeed
explained with the context of $f(R)$ gravity
\cite{Carroll:2003wy}, and the conditions of viable cosmological
models have also been derived \cite{viablemodels}. In the context
of the Solar System regime, severe weak field constraints seem to
rule out most of the models proposed so far
\cite{solartests,Olmo07}, although viable models do exist
\cite{solartests2}.

In the context of dark matter, the possibility that the galactic
dynamics of massive test particles may be understood without the
need for dark matter was also considered in the framework of
$f(R)$ gravity models
\cite{Cap2,Borowiec:2006qr,Mar1,Boehmer:2007kx,Bohmer:2007fh}, and
connections with MOND and the pioneer anomaly further explored by
considering an explicit coupling of an arbitrary function of $R$
with the matter Lagrangian density
\cite{Bertolami:2007gv,Bertolami:2007vu}. The issue of dark matter
is a long outstanding problem in modern astrophysics. Two
observational aspects, namely, the behavior of the galactic
rotation curves and the mass discrepancy in clusters of galaxies
led to the necessity of considering the existence of dark matter
at a galactic and extra-galactic scale. The rotation curves of
spiral galaxies show that the rotational velocities increase from
the center of the galaxy and then attain an approximately constant
value, $v_{tg\infty }\sim 200-300$ km/s, within a distance $r$
from the center of the galaxy~\cite{Bi87}. In these regions the
mass increases linearly with the radius, even where very little
luminous matter can be detected. Relatively to the mass
discrepancy in clusters of galaxies, the total mass of a cluster
can be estimated in two ways. First, by taking into account the
motions of its member galaxies, the virial theorem provides an
estimate, $M_{V}$. Second, the total baryonic mass $M$ may be
estimated by considering the total sum of each individual member's
mass. The mass discrepancy arises as one generally verifies that
$M_{V}$ is considerably greater than $M$, with typical values of
$M_{V}/M\sim 20-30$~\cite{Bi87}. This is usually explained by
postulating the existence of a dark matter, assumed to be a cold
pressure-less medium distributed in a spherical halo around the
galaxies.

Still in the context of modified gravity, an interesting
possibility is the existence of extra dimensions. It is widely
believed that string theory is moving towards a viable quantum
gravity theory, and one of the key predictions of string theory is
precisely the existence of extra spatial dimensions. In the
brane-world scenario, motivated by recent developments in string
theory, the observed 3-dimensional universe is embedded in a
higher-dimensional spacetime~\cite{Maartens1}. Most brane-world
models, including those of the Randall-Sundrum type
\cite{Randall}, produce ultra-violet modifications to General
Relativity, with extra-dimensional gravity dominating at high
energies. However it is also possible for extra-dimensional
gravity to dominate at low energies, leading to infra-red
modifications of General Relativity. New features emerge in the
brane scenario that may be more successful in providing a
covariant infra-red modification of General Relativity, where it
is possible for extra-dimensional gravity to dominate at low
energies.

The Dvali-Gabadadze-Porrati (DGP) models \cite{DGP} achieve this
via a brane induced gravity effect. The generalization of the DGP
models to cosmology lead to late-accelerating cosmologies
\cite{Deffayet}, even in the absence of a dark energy field
\cite{Maartens3}. This exciting feature of ``self acceleration''
may help towards a new resolution to the dark energy problem,
although this model deserves further investigation as a viable
cosmological model \cite{Lue}. While the DGP braneworld offers an
alternative explanation to the standard cosmological model, for
the expansion history of the universe, it offers a paradigm for
nature fundamentally distinct from dark energy models of cosmic
acceleration, even those that perfectly mimic the same expansion
history. It is also fundamental to understand how one may
differentiate this modified theory of gravity from dark energy
models. The DGP braneworld theory also alters the gravitational
interaction itself, yielding unexpected phenomenological
extensions beyond the expansion history. Tests from the solar
system, large scale structure, lensing all offer a window into
understanding the perplexing nature of the cosmic acceleration
and, perhaps, of gravity itself \cite{Lue}. The structure
formation \cite{Maartens4} and the inclusion of inflation are also
important requirements of DGP gravity, if it is to be a realistic
alternative to the standard cosmological model. Generalizations of
the DGP model with the inclusion of a Gauss-Bonnet (GB) term have
also been explored \cite{Maartens3}, the global structure of the
DGP cosmologies have been analyzed \cite{Lue2}, and research into
other brane-world approaches to dark energy, such as the
supersymmetric large extra dimensions (SLED) model have been
undertaken \cite{Burgess}.

In this work, we review several modified theories of gravity,
exploring some of their interesting properties and
characteristics, in particular, focussing mainly on the late-time
cosmic acceleration. We refer the reader to excellent reviews, for
instance, on dark energy and modified gravity in Ref.
\cite{Copeland}, $f(R)$ gravity in Ref. \cite{Sotiriou:2008rp}, an
introduction to modified gravity as an alternative for dark energy
in Ref. \cite{modGB1}, and the late-time acceleration in
braneworlds in Ref. \cite{Koyama:2007rx}. This paper is outlined
in the following manner. In Section \ref{sec:II}, we review $f(R)$
modified theories of gravity, for instance, focussing on the
scalar-tensor representation of $f(R)$ and on the late-time
acceleration; we further analyze an interesting extension to
$f(R)$ gravity by considering an $R-$matter coupling, which yields
some intriguing properties; we also discuss Gauss-Bonnet gravity
and the late-time acceleration, and briefly consider modified
Gauss-Bonnet gravity; the DGP brane-world model is also briefly
reviewed. In Section \ref{sec:III}, we consider the possibility of
dark matter being a geometric effect in $f(R)$ gravity, focussing
on the $f(R)$ generalized virial theorem, and its applications to
galactic cluster observations. Finally in Section \ref{sec:IV}, we
conclude with a summary and discussions.

\section{Modified theories of gravity: Late-time cosmic acceleration}
\label{sec:II}

\subsection{$f(R)$ modified theories of gravity}\label{sec:IIa}

A promising avenue that has been extensively investigated recently
are the $f(R)$ modified theories of gravity, where the standard
Einstein-Hilbert action is replaced by an arbitrary function of
the Ricci scalar $R$~\cite{first}. In this work, we use the metric
formalism, which consists in varying the action with respect to
$g^{\mu\nu}$, although other alternative approaches have been
considered in the literature, namely, the Palatini
formalism~\cite{Palatini,Sotiriou:2006qn}, where the metric and
the connections are treated as separate variables; and the
metric-affine formalism, where the matter part of the action now
depends and is varied with respect to the
connection~\cite{Sotiriou:2006qn}.

\subsubsection{Action and field equations}

The action for the $f(R)$ modified theories of gravity is given by
\begin{equation}
S=\frac{1}{2\kappa}\int d^4x\sqrt{-g}\;f(R)+S_M(g^{\mu\nu},\psi)
\,,
\end{equation}
where $\kappa =8\pi G$. $S_M(g^{\mu\nu},\psi)$ is the matter
action, defined as $S_M=\int d^4x\sqrt{-g}\;{\cal
L}_m(g_{\mu\nu},\psi)$, where ${\cal L}_m$ is the matter
Lagrangian density, in which matter is minimally coupled to the
metric $g_{\mu\nu}$ and $\psi$ collectively denotes the matter
fields.

Using the metric approach, by varying the action with respect to
$g^{\mu\nu}$, provides the following field equation
\begin{equation}
FR_{\mu\nu}-\frac{1}{2}f\,g_{\mu\nu}-\nabla_\mu \nabla_\nu
F+g_{\mu\nu}\Box F=\kappa\,T^{(m)}_{\mu\nu} \,,
    \label{field:eq}
\end{equation}
where $F\equiv df/dR$. The matter stress-energy tensor, $T_{\mu
\nu}^{(m)}$, is defined as
\begin{equation}
T_{\mu \nu
}^{(m)}=-\frac{2}{\sqrt{-g}}\frac{\delta(\sqrt{-g}\,{\cal
L}_m)}{\delta(g^{\mu\nu})} ~.
 \label{defSET}
\end{equation}

Now, considering the contraction of Eq. (\ref{field:eq}), provides
the following relationship
\begin{equation}
FR-2f+3\,\Box F=\kappa\,T \,,
 \label{trace}
\end{equation}
which shows that the Ricci scalar is a fully dynamical degree of
freedom.

Note that the field equation, Eq. (\ref{field:eq}), may be written
as
\begin{equation}
G_{\mu\nu}\equiv R_{\mu\nu}-\frac{1}{2}R\,g_{\mu\nu}=\kappa
T^{{\rm eff}}_{\mu\nu} \,,
    \label{field:eq2}
\end{equation}
where the effective stress energy tensor is given by $T^{{\rm
eff}}_{\mu\nu}= T^{(c)}_{\mu\nu}+\tilde{T}^{(m)}_{\mu\nu}$. The
components $\tilde{T}^{(m)}_{\mu\nu}$ and the curvature
stress-energy tensor, $T^{(c)}_{\mu\nu}$, are defined as
\begin{eqnarray}
\tilde{T}^{(m)}_{\mu\nu}&=&T^{(m)}_{\mu\nu}/F \,, \\
T^{(c)}_{\mu\nu}&=&\frac{1}{\kappa F}\left[\nabla_\mu \nabla_\nu F
-\frac{1}{4}g_{\mu\nu}\left(RF+\Box F+\kappa T\right) \right] \,,
    \label{gravfluid}
\end{eqnarray}
respectively. It is also interesting to consider the conservation
law for the above curvature stress-energy tensor. Taking into
account the Bianchi identities, $\nabla^\mu G_{\mu\nu}=0$, and the
diffeomorphism invariance of the matter part of the action, which
yields $\nabla^\mu T^{(m)}_{\mu\nu}=0$, we verify that the
effective Einstein field equation provides the following
conservation law
\begin{equation}\label{conserv-law}
\nabla^\mu T^{(c)}_{\mu\nu}=\frac{1}{F^2}
T^{(m)}_{\mu\nu}\nabla^\mu F  \,.
\end{equation}

\subsubsection{Scalar-tensor representation for $f(R)$ gravity}
\label{sec:scalar-tensor}

$f(R)$ gravity may be written as a scalar-tensor theory, by
introducing a Legendre transformation $\left\{ R,f\right\}
\rightarrow \left\{ \phi ,V\right\} $ defined as
\begin{equation}
\phi \equiv F\left( R\right), \quad V\left( \phi \right) \equiv
R\left( \phi \right) F-f\left( R\left( \phi \right) \right) \,.
\end{equation}
In this representation the field equations of $f(R)$ gravity can
be derived from a Brans-Dicke type action with parameter $\omega
=0$, given by
\begin{equation}
S=\frac{1}{2\kappa}\int \left[ \phi R-V(\phi) +L_{m}\right]
\sqrt{-g}\; d^{4}x.
\end{equation}
The only requirement for the gravitational field equations to be
expressed in the form of a Brans-Dicke theory is that $F(R)$ be
invertible, that is, $R(F)$ exists~\cite{Olmo07}.

Thus, the field equations of $f(R)$ gravity can be reformulated as
\begin{equation}
R_{\mu \nu }-\frac{1}{2}g_{\mu \nu }R=8\pi \frac{G}{\phi }T_{\mu
\nu }+\theta _{\mu \nu },  \label{fieldin}
\end{equation}
where
\begin{equation}
\theta _{\mu \nu }=-\frac{1}{2}V\left( \phi \right) g_{\mu \nu }+\frac{1}{%
\phi }\left( \nabla _{\mu }\nabla _{\nu }-g_{\mu \nu }\square
\right) \phi .
\end{equation}
Using these variables the trace equation, i.e., Eq.~(\ref{trace}),
takes the form
\begin{equation}  \label{trace1}
3 \square \phi + 2V(\phi) -\phi \frac{dV}{d\phi }=\kappa T.
\end{equation}

The modification of the standard Einstein-Hilbert action leads to
the appearance of an effective gravitational constant
$G_{\mathrm{eff}}=G/\phi$ in the field equations. Note the
presence of a new effective source term for the gravitational
field, given by the tensor $\theta_{\mu \nu}$.

\subsubsection{$R-$matter couplings in $f(R)$ gravity}

Recently, in the context of $f(R)$ theories of modified gravity,
it was shown that a function of $R-$matter coupling induces a
non-vanishing covariant derivative of the stress-energy,
$\nabla_\mu T^{\mu\nu} \neq 0$. This potentially leads to a
deviation from geodesic motion, and consequently the appearance of
an extra force \cite{Bertolami:2007gv}. Implications, for
instance, for stellar equilibrium have been studied in Ref.
\cite{Bertolami:2007vu}. The equivalence with scalar-tensor
theories with two scalar fields has been considered in Ref.
\cite{Bertolami:2008im}, and a viability stability criterion was
also analyzed in Ref. \cite{Faraoni:2007sn}. It is interesting to
note that nonlinear couplings of matter with gravity were analyzed
in the context of the accelerated expansion of the Universe
\cite{Odintsov}, and in the study of the cosmological constant
problem \cite{Lambda}.

The action for $R-$matter couplings, in $f(R)$ modified theories
of gravity \cite{Bertolami:2007gv}, takes the following form
\begin{equation}
S=\int \left\{\frac{1}{2}f_1(R)+\left[1+\lambda f_2(R)\right]{\cal
L}_{m}\right\} \sqrt{-g}\;d^{4}x~,
\end{equation}
where $f_i(R)$ (with $i=1,2$) are arbitrary functions of the
curvature scalar $R$. For notational simplicity we consider
$\kappa=1$ throughout this subsection.

Varying the action with respect to the metric $g^{\mu\nu}$ yields
the field equations, given by
\begin{equation}
F_1R_{\mu \nu }-\frac{1}{2}f_1g_{\mu \nu }-\nabla_\mu \nabla_\nu
\,F_1+g_{\mu\nu}\square F_1=-2\lambda F_2{\cal L}_m R_{\mu\nu}
+2\lambda(\nabla_\mu \nabla_\nu-g_{\mu\nu}\square){\cal L}_m F_2
+(1+\lambda f_2)T_{\mu \nu }^{(m)}\,, \label{field}
\end{equation}
where we have denoted $F_i(R)=f'_i(R)$, and the prime represents
the derivative with respect to the scalar curvature.

Now, taking into account the generalized Bianchi
identities~\cite{Bertolami:2007gv,Koivisto}, one deduces the
following corrected conservation equation
\begin{equation}
\nabla^\mu T_{\mu \nu }^{(m)}=\frac{\lambda F_2}{1+\lambda
f_2}\left[g_{\mu\nu}{\cal L}_m- T_{\mu \nu
}^{(m)}\right]\nabla^\mu R ~, \label{cons1}
\end{equation}
where the coupling between the matter and the higher derivative
curvature terms describes an exchange of energy and momentum
between both.

In the following, consider the equation of state for a perfect
fluid
\begin{equation}
T_{\mu\nu}^{(m)}=\left(\rho +p\right)U_{\mu}U_{\nu}+pg_{\mu\nu}
\,,
   \label{perfectfluid}
\end{equation}
where $\rho$ is the energy density and $p$, the pressure,
respectively. The four-velocity, $U_{\mu }$, satisfies the
conditions $U_{\mu }U^{\mu }=-1$ and $U^{\mu }U_{\mu ;\nu }=0$.

Introducing the projection operator
$h_{\mu\nu}=g_{\mu\nu}+U_{\mu}U_{\nu}$, gives rise to non-geodesic
motion governed by the following equation of motion for a fluid
element: $dU^{\mu}/ds+\Gamma _{\alpha
\beta}^{\mu}U^{\alpha}U^{\beta}=f^{\mu}$, where the extra force,
$f^{\mu}$, is given by
\begin{equation}
f^{\mu}=\frac{1}{\rho +p}\left[\frac{\lambda F_2}{1+\lambda
f_2}\left({\cal L}_m-p\right)\nabla_\nu R+\nabla_\nu p \right]
h^{\mu \nu }\,.
     \label{force}
\end{equation}

In a recent paper \cite{Sotiriou:2008it}, the authors argued that
a ``natural choice'' for the matter Lagrangian density for perfect
fluids is ${\cal L}_m=p$, based on Refs.
\cite{Schutz:1970my,Brown:1992kc}, where $p$ is the pressure. This
choice has a particularly interesting application in the analysis
of the $R-$matter coupling for perfect fluids, which implies in
the vanishing of the extra force \cite{Bertolami:2007gv}. However,
it is important to point out that despite the fact that ${\cal
L}_m=p$ does indeed reproduce the perfect fluid equation of state,
it is not unique \cite{BLP}. Other choices include, for instance,
${\cal L}_m=-\rho$ \cite{Brown:1992kc,HawkingEllis}, where $\rho$
is the energy density, or ${\cal L}_m=-na$, were $n$ is the
particle number density, and $a$ is the physical free energy
defined as $a=\rho/n-Ts$, with $T$ being the temperature and $s$
the entropy per particle (see Ref. \cite{BLP,Brown:1992kc} for
details).

Hence, it is clear that no immediate conclusion may be extracted
regarding the additional force imposed by the non-minimum coupling
of curvature to matter, given the different available choices for
the Lagrangian density. One may conjecture that there is a deeper
principle or symmetry that provides a unique Lagrangian density
for a perfect fluid \cite{BLP}. This has not been given due
attention in the literature, as arbitrary gravitational field
equations depending on the matter Lagrangian have not always been
the object of close analysis. See Ref. \cite{BLP} for more
details.

\subsubsection{Late-time cosmic acceleration}

In this subsection, we show that $f(R)$ gravity may lead to an
effective dark energy, without the need to introduce a negative
pressure ideal fluid. Consider the FLRW metric
\begin{equation}
ds^2=-dt^2+a^2(t)\left[\frac{dr^2}{1-kr^2}
+r^2(d\theta^2+\sin^2\theta \,d\phi^2)\right] \,.
\end{equation}
Taking into account the perfect fluid description for matter given
by Eq. (\ref{perfectfluid}), we verify that the gravitational
field equation, Eq. (\ref{field:eq2}), provides the generalised
Friedmann equations in the following form
\cite{Capozziello:2003tk,Sotiriou:2007yd}:
\begin{eqnarray}
\left(\frac{\dot{a}}{a}\right)^2-\frac{1}{3F(R)}\left\{\frac{1}{2}
\left[f(R)-RF(R)\right]-3\left(\frac{\dot{a}}{a}\right)\dot{R}
F'(R)\right\}&=&\frac{\kappa}{3}\rho \,,  \\
\left(\frac{\ddot{a}}{a}\right)+\frac{1}{2F(R)}\left\{\frac{\dot{a}}{a}\dot{R}
F'(R)+\ddot{R}F'(R)+\dot{R}^2F''(R)-\frac{1}{3}\left[f(R)-RF(R)\right]\right\}
&=&-\frac{\kappa}{6}(\rho+3p) \,.
\end{eqnarray}

These modified Friedmann field equations may be rewritten in a
more familiar form, as
\begin{eqnarray}
\left(\frac{\dot{a}}{a}\right)^2&=&\frac{\kappa}{3}\rho_{\rm tot} \,,  \\
\left(\frac{\ddot{a}}{a}\right)&=&-\frac{\kappa}{6}(\rho_{\rm
tot}+3p_{\rm tot}) \,,
    \label{rho+3p}
\end{eqnarray}
where $\rho_{\rm tot}=\rho+\rho_{(c)}$ and $p_{\rm
tot}=p+p_{(c)}$, and the curvature stress-energy components,
$\rho_{(c)}$ and $p_{(c)}$, are defined as
\begin{eqnarray}
\rho_{(c)}&=&\frac{1}{\kappa F(R)}\left\{\frac{1}{2}
\left[f(R)-RF(R)\right]-3\left(\frac{\dot{a}}{a}\right)\dot{R}
F'(R)\right\} \,,  \\
p_{(c)}&=&\frac{1}{\kappa
F(R)}\left\{2\left(\frac{\dot{a}}{a}\right)\dot{R}
F'(R)+\ddot{R}F'(R)+\dot{R}^2F''(R)-\frac{1}{2}\left[f(R)-RF(R)\right]\right\}
 \,,
\end{eqnarray}
respectively. The late-time cosmic acceleration is achieved if the
condition $\rho_{\rm tot}+3p_{\rm tot}<0$ is obeyed, which follows
from Eq. (\ref{rho+3p}).

For simplicity, consider the absence of matter, $\rho=p=0$. Now,
taking into account the equation of state $\omega_{\rm
eff}=p_{(c)}/\rho_{(c)}$, with $f(R)\propto R^n$ and a generic
power law $a(t)=a_0(t/t_0)^\alpha$ \cite{Capozziello:2003tk}, the
parameters $\omega_{\rm eff}$ and $\alpha$ are given by
\begin{equation}
\omega_{\rm eff}=-\frac{6n^2-7n-1}{6n^2-9n+3} \,, \qquad
\alpha=\frac{-2n^2+3n-1}{n-2} \,,
\end{equation}
respectively, for $n\neq 1$. Note that a suitable choice of $n$
can lead to the desired value of $\omega_{\rm eff}<-1/3$,
achieving the late-time cosmic acceleration.

Another example is a model of the form $f(R)=R-\mu^{2(n+1)}/R^n$
\cite{Carroll:2003wy}. Choosing once again a generic power law for
the scale factor, the parameter can be written as
\begin{equation}
\omega_{\rm eff}=-1+\frac{2(n+2)}{3(2n+1)(n+1)} \,,
\end{equation}
and once again a desired value of $\omega_{\rm eff}<-1/3$ may be
attained, by appropriately choosing the value of the parameter
$n$. Note that as $n\rightarrow \infty$ the spacetime is
approximately de Sitter.

Other forms of $f(R)$ have also been considered in the literature,
for instance those involving logarithmic terms, such as
$f(R)=R+\alpha \ln (R/\mu^2)+\beta R^m$ or $f(R)=R+\gamma R^{-n}
[\ln (R/\mu^2)]^m$ \cite{modGB1,Nojiri:2003ni}. These models also
yield acceptable values for the effective equation of state
parameter, resulting in the late-time cosmic acceleration.

\subsection{Gauss-Bonnet gravity and cosmic acceleration}\label{sec:IIb}

\subsubsection{Gauss-Bonnet gravity}

In considering alternative higher-order gravity theories, one is
liable to be motivated in pursuing models consistent and inspired
by several candidates of a fundamental theory of quantum gravity.
In this context, it may be possible that unusual gravity-matter
couplings predicted by string/M-theory may become important at the
recent low-curvature Universe. For instance, one may couple a
scalar field not only with the curvature scalar, as in
scalar-tensor theories, but also with higher order curvature
invariants. Indeed, motivations from string/M-theory predict that
scalar field couplings with the Gauss-Bonnet invariant ${\cal G}$
are important in the appearance of non-singular early time
cosmologies. It is also possible to apply these motivations to the
late-time Universe in an effective Gauss-Bonnet dark energy model
\cite{Nojiri:2005vv}.

Consider the action of Gauss-Bonnet gravity given by
\begin{equation}
S=\int d^4x
\sqrt{-g}\left[\frac{R}{2\kappa}-\frac{\lambda}{2}\partial_\mu
\phi \partial^\mu\phi-V(\phi)+f(\phi){\cal
G}\right]+S_M(g^{\mu\nu},\psi)\,,
    \label{GBaction}
\end{equation}
where $\lambda=+1$ is defined for a canonical scalar field, and
$\lambda=-1$ for a phantom field, respectively. The Gauss-Bonnet
invariant in given by ${\cal G}\equiv
R^2-4R_{\mu\nu}R^{\mu\nu}+R_{\mu\nu\alpha\beta}R^{\mu\nu\alpha\beta}$.
It is also important to note that in the matter action, matter is
minimally coupled to the metric and not to the scalar field,
making Gauss-Bonnet gravity a metric theory. Thus, using the
diffeomorphism invariance of $S_M(g^{\mu\nu},\psi)$ yields the
covariant conservation of the stress-energy tensor, $\nabla^\mu
T_{\mu\nu}^{(m)}=0$.

Varying the action with respect to $\phi$, provides the equation
of motion for the scalar field, given by
\begin{equation}
\lambda\nabla^2\phi-V'(\phi)+f'(\phi){\cal G}=0\,.
\end{equation}

The gravitational field equations are given by varying the action
with respect to the metric $g^{\mu\nu}$, and provides the
following relationship \cite{Nojiri:2005vv,Sotiriou:2007yd}
\begin{eqnarray}
&&\frac{1}{\kappa}G^{\mu\nu}-\frac{1}{2}g^{\mu\nu}f(\phi){\cal G}
+2f(\phi)RR^{\mu\nu}-2\nabla^\mu
\nabla^\nu[f(\phi)R]+2g^{\mu\nu}\nabla^2[f(\phi)R]
    \nonumber   \\
&&-8f(\phi)R^{\mu}{}_{\rho}R^{\nu\rho}
+4\nabla_\rho\nabla^\mu[f(\phi)R^{\nu\rho}]
+4\nabla_\rho\nabla^\nu[f(\phi)R^{\mu\rho}]
-4\nabla^2[f(\phi)R^{\mu\nu}]
    \nonumber   \\
&&-4g^{\mu\nu}\nabla_\rho \nabla_\sigma[f(\phi)R^{\rho\sigma}]
+2f(\phi)R^{\mu\rho\sigma\tau} R^{\nu}{}_{\rho\sigma\tau}-
4\nabla_\rho\nabla_\sigma [f(\phi)R^{\mu\rho\sigma\nu}]
=T^{\mu\nu}+T_{\phi}^{\mu\nu}  \,,
    \label{GBgravFEq}
\end{eqnarray}
where $T_{\phi}^{\mu\nu}$ is given by
\begin{equation}
T_{\phi}^{\mu\nu}=\lambda\left(\frac{1}{2}\partial^\mu\phi
\;\partial^\nu\phi-\frac{1}{4}g^{\mu\nu}\partial_\rho\phi
\;\partial^\rho\phi\right) -\frac{1}{2}g^{\mu\nu}V(\phi)\,.
\end{equation}

One may use the following identities obtained from the Bianchi
identity \cite{Nojiri:2005vv,Sotiriou:2007yd}:
\begin{eqnarray}
&&\nabla^\rho R_{\rho\tau\mu\nu}=\nabla_\mu R_{\nu\tau}-\nabla_\nu
R_{\mu\tau}\,, \qquad \nabla^\nu R_{\nu\mu}=\frac{1}{2}\nabla_\mu
R\,, \qquad  \nabla_\mu\nabla_\nu R^{\mu\nu}=\frac{1}{2}\Box R\,,
     \nonumber   \\
&&\nabla_\rho\nabla_\sigma
R^{\mu\rho\nu\sigma}=\nabla^2R^{\mu\nu}-\frac{1}{2}\nabla^\mu\nabla^\nu
R+R^{\mu\rho\nu\sigma}R_{\rho\sigma}-R^{\mu}{}_{\rho}R^{\nu\rho}
\,,
    \nonumber    \\
&&\nabla_\rho\nabla^{(\mu} R^{\nu )\rho}
=\frac{1}{2}\nabla^{(\mu}\nabla^{\nu
)}R-R^{\mu\rho\nu\sigma}R_{\rho\sigma}+R^{\mu}{}_{\rho}R^{\nu\rho}
\,,
   \nonumber \label{GBBianchiId}
\end{eqnarray}
in the Gauss-Bonnet gravitational field equation, which may then
be formally simplified to
\begin{eqnarray}
&&\frac{1}{\kappa}G^{\mu\nu}-\frac{1}{2}g^{\mu\nu}f(\phi){\cal G}
+2f(\phi)RR^{\mu\nu}+4f(\phi)R^{\mu}{}_{\rho}R^{\nu\rho}
    \nonumber   \\
&& +2f(\phi)R^{\mu\rho\sigma\tau} R^{\nu}{}_{\rho\sigma\tau}-
4f(\phi)R^{\mu\rho\sigma\nu} R_{\rho\sigma}
=T^{\mu\nu}+T_{\phi}^{\mu\nu} +T_{c}^{\mu\nu} \,,
    \label{GBgravFEq2}
\end{eqnarray}
where $T_{c}^{\mu\nu}$ is defined as
\begin{eqnarray}
T_{c}^{\mu\nu}&=&2[\nabla^\mu \nabla^\nu
f(\phi)]R-2g^{\mu\nu}[\nabla^2 f(\phi)]R -4[\nabla_\rho\nabla^\mu
f(\phi)]R^{\nu\rho}-4[\nabla_\rho\nabla^\nu f(\phi)]R^{\mu\rho}
      \nonumber   \\
&&+4[\nabla^2f(\phi)]R^{\mu\nu}+4g^{\mu\nu}[\nabla_\rho
\nabla_\sigma f(\phi)]R^{\rho\sigma}- 4[\nabla_\rho\nabla_\sigma
f(\phi)]R^{\mu\rho\sigma\nu} \,.
    \label{GB-SETcurv}
\end{eqnarray}

Using the FLRW metric, the modified Friedmann equations for
Gauss-Bonnet gravity reduce to the following relationships
\begin{eqnarray}
\left(\frac{\dot{a}}{a}\right)^2&=&\frac{1}{3}\kappa
\left(\rho+\frac{\lambda}{2}\dot{\phi}^2+V(\phi)
-24\dot{\phi}f'(\phi)H^3\right) \,,  \label{GBFried1}
        \\
\left(\frac{\ddot{a}}{a}\right)&=&-\frac{\kappa}{6}(\rho+3p)
-\frac{\kappa}{3}\left[\lambda\dot{\phi}^2-V(\phi)
+12H^3\dot{\phi}f'(\phi)+12\frac{\partial}{\partial
t}\left(H^2\dot{f}\right)\right] \,,
    \label{GBFried2}
\end{eqnarray}
and the equation of motion for the scalar field is given by
\begin{equation}
\lambda\left(\ddot{\phi}+3H\dot{\phi}\right)+V'(\phi)
-24f'(\phi)H^2\left(\dot{H}+H^2\right)=0\,,
  \label{GBscalarEOM}
\end{equation}
where the expression for the Gauss-Bonnet invariant, ${\cal
G}=24H^2(\dot{H}+H^2)$, is used.

Note that in the absence of matter, the Gauss-Bonnet gravitational
field equations may be written as
\begin{equation}
\rho_{\rm GB}=\frac{3}{\kappa}H^2\,, \qquad p_{\rm
GB}=-\frac{1}{\kappa}\left(3H^2+2\dot{H}\right)\,,
\end{equation}
where the Gauss-Bonnet curvature stress-energy tensor components
are defined as
\begin{eqnarray}
\rho_{\rm
GB}&=&\frac{\lambda}{2}\dot{\phi}^2+V(\phi)-24\dot{\phi}f'(\phi)H^3
\,, \label{GBrho}   \\
p_{\rm GB}
&=&\frac{\lambda}{2}\dot{\phi}^2-V(\phi)+8\frac{\partial}{\partial
t}\left(H^2\dot{f}\right)+16H^3\dot{\phi} f'(\phi) \,.
    \label{GBp}
\end{eqnarray}
These relationships are particularly interesting as one may now
define an effective equation of state given by
\begin{equation}
\omega_{\rm eff}=\frac{p_{\rm GB}}{\rho_{\rm
GB}}=-1-\frac{2\dot{H}}{3H^2}\,.
    \label{GBeffEOS}
\end{equation}

Consider the choices of an exponential scalar-GB coupling and
exponential scalar potential given by \cite{Nojiri:2005vv}
\begin{equation}
V(\phi)=V_0 e^{-2\phi/\phi_0}\,, \qquad  f(\phi)=f_0
e^{2\phi/\phi_0}\,.
\end{equation}
The scale factor and the scalar field are respectively chosen in
the following form
\begin{equation}
a(t)= \left\{ \begin{array}{ll}
a_0t^{h_0}, \qquad \quad \;\;\,{\rm for} \quad h_0>0 \;\;({\rm quintessence}),\\
a_0(t_s-t)^{h_0}, \quad {\rm for} \quad h_0<0 \;\;({\rm phantom}),
\end{array}
\right.
\end{equation}
and
\begin{equation}
\phi(t)= \left\{ \begin{array}{ll}
\phi_0\ln\left(\frac{t}{t_1}\right), \qquad \;\;\;{\rm for} \quad h_0>0 ,\\
\phi_0\ln\left(\frac{t_s-t}{t_1}\right), \qquad {\rm for} \quad
h_0<0,
\end{array}
\right.
\end{equation}
where $t_1$ is an arbitrary constant~\cite{Nojiri:2005vv}.

Using the above choices, and considering the absence of matter,
then taking into account Eqs. (\ref{GBFried1}) and
(\ref{GBscalarEOM}), and finally reorganizing the terms leads to
the following equations
\begin{eqnarray}
V_0t_1^2&=&-\frac{1}{\kappa(1+h_0)}\left[3h_0^2(1-h_0)
+\frac{\lambda\phi_0^2\kappa(1-5h_0)}{2}\right]
\,, \label{GBFried1c}   \\
\frac{48f_0h_0^2}{t_1^2} &=&-\frac{6}{\kappa(1+h_0)}\left(h_0
-\frac{\lambda\phi_0^2\kappa}{2}\right) \,.
    \label{GBFried2c}
\end{eqnarray}

The effective equation of state parameter, Eq. (\ref{GBeffEOS}),
takes the following form
\begin{equation}
\omega_{\rm eff}=-1-\frac{2}{3h_0}\,.
    \label{GBeffEOS2}
\end{equation}
Note that if $h_0<0$ then $\omega_{\rm eff}<-1$, reflecting an
effective phantom regime; and if $h_0>0$ then $\omega_{\rm
eff}>-1$, which reflects an effective quintessence regime.
However, it has been shown that the case of $h_0<0$ is always
stable, while the case of a non-phantom $h_0>0$ cosmology is
always unstable \cite{Nojiri:2005vv}. An interesting example is
the case of $V_0=0$, which from Eq. (\ref{GBFried1c}) imposes the
following condition \cite{Nojiri:2005vv}:
$\phi_0^2=-6h_0^2(1-h_0)/[\lambda\kappa(1-5h_0)]$. In order for
$\phi_0$ to be real, in the case of the a canonical scalar,
$\lambda=1$, one finds that $1/5<h_0<1$; for a phantom field,
$\lambda=-1$, then $h_0<1/5$ or $h_0\geq 1$. To achieve an
effective equation of state parameter value that mimics dark
energy is to consider, for instance, $h_0=-80/3$, which imposes
the $\omega_{\rm eff}=-1.025$.

We refer the reader to Ref. \cite{Nojiri:2005vv} for more examples
and details. It is also interesting to note that observational
constraints \cite{dataGB}, such as CMBR, galaxy distribution,
large scale structure and supernovae seem to favour the
Gauss-Bonnet coupling.

\subsubsection{Modified Gauss-Bonnet gravity and the late-time
acceleration}

An interesting alternative gravitational theory is modified
Gauss-Bonnet gravity, which is given by the following action:
\begin{equation}
S=\int d^4x \sqrt{-g}\left[\frac{R}{2\kappa}+f({\cal
G})\right]+S_M(g^{\mu\nu},\psi)\,.
   \label{modGBaction}
\end{equation}
This theory has been extensively analyzed in the literature
\cite{modGB1,modGB2,Nojiri:2005jg}, and rather than review all of
its intricate details here, we note that it is a subset of
Gauss-Bonnet gravity given by the action (\ref{GBaction}).

To see this, we follow closely the approach outlined in Ref.
\cite{Sotiriou:2007yd,Nojiri:2005jg}. By introducing two auxiliary
scalar fields $A$ and $B$, the gravitational part of the action
(\ref{modGBaction}), may be rewritten as
\begin{equation}
S=\int d^4x \sqrt{-g}\left[\frac{R}{2\kappa}+B({\cal
G}-A)+f(A)\right]\,.
   \label{modGBaction2}
\end{equation}
Now varying with respect to $B$, one obtains $A={\cal G}$, so that
the action (\ref{modGBaction2}) is recovered. Varying with respect
to $A$, one obtains $B=f'(A)$, and substituting in
(\ref{modGBaction2}) leads to
\begin{equation}
S=\int d^4x \sqrt{-g}\left[\frac{R}{2\kappa}+f'(A)({\cal
G}-A)+f(A)\right]\,.
   \label{modGBaction3}
\end{equation}

With the following definitions
\begin{equation}
\phi=A\,, \qquad  {\rm and} \qquad V(\phi)=Af'(A)-f(A)\,,
\end{equation}
one finally ends up with the following action
\begin{equation}
S=\int d^4x \sqrt{-g}\left[\frac{R}{2\kappa}-V(\phi)+f(\phi){\cal
G}\right]+S_M(g^{\mu\nu},\psi)\,.
    \label{GBaction4}
\end{equation}
which is simply the action for Gauss-Bonnet gravity given by Eq.
(\ref{GBaction}) with the absence of the kinetic term. Thus,
modified Gauss-Bonnet theory, given by (\ref{modGBaction}) is
dynamically equivalent to Gauss-Bonnet gravity with $\lambda=0$
\cite{Sotiriou:2007yd,Nojiri:2005jg}. We refer the reader to Refs.
\cite{modGB1,modGB2} for more details.

\subsection{DGP brane gravity and self-acceleration}\label{sec:IIc}

One of the key predictions of string theory is the existence of
extra spatial dimensions. In the brane-world scenario, motivated
by recent developments in string theory, the observed
3-dimensional universe is embedded in a higher-dimensional
spacetime~\cite{Maartens1}. One of the simplest covariant models
is the Dvali-Gabadadze-Porrati (DGP) braneworld model, in which
gravity leaks off the $4D$ Minkowski  brane into the $5D$ bulk at
large scales. The generalization of the DGP models to cosmology
lead to late-accelerating cosmologies \cite{Deffayet}, even in the
absence of a dark energy field \cite{Maartens3}. While the DGP
braneworld offers an alternative explanation to the standard
cosmological model, for the expansion history of the universe, it
offers a paradigm for nature fundamentally distinct from dark
energy models of cosmic acceleration, even those that perfectly
mimic the same expansion history.

The $5D$ action describing the DGP model is given by
\begin{equation}
S=\frac{1}{2\kappa_5}\int
d^5x\sqrt{-g}\;^{(5)}R+\frac{1}{2\kappa_4}\int
d^4x\sqrt{-\gamma}\;^{(4)}R-\int d^4x\sqrt{-\gamma}\;{\cal L}_m
\,,
\end{equation}
where $\kappa_5=8\pi G_5$ and $\kappa_4=8\pi G_4$. The first term
in the action is the Einstein-Hilbert action in five dimensions
for a five-dimensional bulk metric, $g_{AB}$, with a
five-dimensional Ricci scalar $^{(5)}R$; the second term is the
induced Einstein-Hilbert term on the brane, with a
four-dimensional induced metric $\gamma$ on the brane; and ${\cal
L}_m$ represents the matter Lagrangian density confined to the
brane.

The transition from $4D$ to $5D$ behavior is governed by a
cross-over scale, $r_c$, given by
\begin{equation}
r_c=\frac{\kappa_5}{2\kappa_4} \,.
\end{equation}
Gravity manifests itself as a $4$-dimensional theory for
characteristic scales much smaller than $r_c$; for large distances
compared to $r_c$, one verifies a leakage of gravity into the
bulk, consequently making the higher dimensional effects
important. Thus, the leakage of gravity at late times initiates
acceleration, due to the weakening of gravity on the brane.

For weak fields, the gravitational potential behaves as
\begin{equation}
\Phi \sim \left\{ \begin{array}{ll}
r^{-1}, \quad {\rm for} \quad r<r_c,\\
r^{-2}, \quad {\rm for}\quad r>r_c.
\end{array}
\right.
\end{equation}

Taking into account the FRW metric, and considering a flat
geometry, the modified Friedmann equation is given by
\begin{equation}
H^2-\frac{\epsilon}{r_c}H=\frac{8\pi G}{3}\rho\,,
   \label{DGPfriedeq}
\end{equation}
where $\epsilon=\pm 1$, and the energy density satisfies the
standard conservation equation, i.e., $\dot{\rho}+3H(\rho+p)=0$.
For scales $H^{-1}\ll r_c$, the second term is negligible, and Eq.
(\ref{DGPfriedeq}) reduces to the general relativistic Friedmann
equation, i.e., $H^2=8\pi G\rho/3$. The second term becomes
significant for scales comparable to the cross-over scale,
$H^{-1}\geq r_c$. Self-acceleration occurs for the branch
$\epsilon =+1$, and the modified Friedmann equation shows that at
late times in a CDM universe characterized by a scale factor $\rho
\propto a^{-3}$, the universe approaches a de Sitter solution
\begin{equation}
H\rightarrow H_{\infty}=\frac{1}{r_c}
\end{equation}
Thus, one may achieve late-time acceleration if the $H_0$ is of
the order of $H_{\infty}$. Note that the late-time acceleration in
the DGP model is not due to the presence of a negative pressure,
but simply due to the weakening of gravity on the brane as a
consequence of gravity leakage at late times.

Although the weak-field gravitational DGP behaves as $4D$ on
scales smaller than $r_c$, linearized DGP gravity is not described
by General Relativity \cite{Koyama:2007rx,Koyama:2007za}. It is
also interesting to note that despite the fact that the expansion
history of the DGP model and General Relativity are the same, the
structure formation in both are essentially different
\cite{Koyama:2005kd}. Combining these features provides the
possibility of distinguishing the DGP model from dark energy
models in General Relativity. We should also emphasize that while
the DGP model is not ruled out by current observations, the
$\Lambda$CDM model fits the data comfortably \cite{dataDGP}.
Another interesting aspect of this model is that the
self-accelerating branch in the DGP model contains a ghost at the
linearized level \cite{Koyama:2005tx,Koyama:2007za}. The presence
of the ghost implies a negative sign for the kinetic term,
resulting in negative energy densities, consequently leading to
the instability of the spacetime. However, in a recent paper it
was claimed that a higher codimension generalization of the DGP
scenario is free of ghost instabilities \cite{deRham:2007xp}, and
further work along these lines is currently underway. We refer the
reader to Ref. \cite{Koyama:2007rx,Koyama:2007za}, and references
therein, for more details on the DGP model.

\section{Dark matter as a geometric effect of modified gravity}
\label{sec:III}

The issue of dark matter is a long outstanding problem in modern
astrophysics. Two observational aspects, namely, the behavior of
the galactic rotation curves and the mass discrepancy in clusters
of galaxies led to the necessity of considering the existence of
dark matter at a galactic and extra-galactic scale. The galactic
rotation curves of spiral galaxies~\cite{Bi87} are probably the
most striking evidences for the possible failure of Newtonian
gravity and of General Relativity on galactic and intergalactic
scales. In these galaxies, neutral hydrogen clouds are observed at
large distances from the center, much beyond the extent of the
luminous matter. As these clouds are moving in circular orbits
with nearly constant tangential velocity $v_{\mathrm{tg}}$, such
orbits are maintained by the balance between the centrifugal
acceleration $v_{\mathrm{tg}}^{2}/r$ and the gravitational
attraction $GM(r)/r^{2}$ of the total mass $M(r)$ contained within
the radius $r$. This yields an expression for the galactic mass
profile of the form $M(r)=rv_{\mathrm{tg}}^{2}/G$, with the mass
increasing linearly with $r $, even at large distances, where very
little luminous matter has been detected~\cite{Bi87}. This
peculiar behavior of the rotation curves is usually explained by
postulating the existence of dark matter, assumed to be a cold and
pressureless medium, distributed in a spherical halo around the
galaxies. There are many possible candidates for dark matter, the
most popular ones being the weakly interacting massive particles
(WIMP)~\cite{OvWe04}.

One cannot also \textit{a priori} exclude the possibility that
Einstein's (and Newton's) theory of gravity breaks down at
galactic scales. In this context, several theoretical models,
based on a modification of Newton's law or of General Relativity,
have been proposed to explain the behavior of the galactic
rotation curves~\cite{expl}. A promising avenue that has been
extensively investigated recently are the $f(R)$ modified theories
of gravity. In this context, early work in explaining dark matter
using $f(R)$ gravity using models of the form $f(R)\propto R^n$
found large values for $n$ \cite{Cap2,Borowiec:2006qr,Mar1}.

In these papers, a power law modified Newtonian potential of the
form $\Phi(r) = -\frac{G m}{2r} [1+(r/r_c)^{\beta}]$ was
considered, to describe the observed behavior of the galactic
rotation curves, where $m$ is the mass of the particle, $r_c$ a
constant and the coefficient $\beta$ depends on the `slope'
parameter $n$ in the modified action. Using this modified
Newtonian potential, it was found that the best fit to $15$ low
luminosity rotation curves in $R^n $ gravity is obtained for
$n=3.5$~\cite{Cap2} (somewhat lower values, in particular,
$n=2.2$, were obtained in~\cite{Borowiec:2006qr,Mar1}). These
results seem to suggest that a strong modification of standard
general relativity is required to explain the observed behavior of
the galactic rotation curves. Note that these large values of $n$
are in gross violation with the Solar System tests.

However, recently is was found that only slight deviations from
General Relativity are needed, i.e., $n=1+\epsilon$ with $\epsilon
\ll 1$ \cite{Boehmer:2007kx,Bohmer:2007fh}. This discrepancy of
values can be traced back to the correction term of the modified
Newtonian potential. It was shown in Ref. \cite{Boehmer:2007kx}
that the correct modified term to the Newtonian potential in the
``dark matter'' dominated region, where the rotation curves are
strictly flat, must have a logarithmic dependence on the radial
coordinate $r$, of the form $\Phi_{N}(r)= -\frac{G m}{2r}+ v_{\rm
tg}^2\ln(r/r_{0})$, where $r_0$ is an arbitrary constant of
integration (we refer the reader to Ref. \cite{Boehmer:2007kx} for
details). These differences in the Newtonian limit in the two
models result in different values of the parameter $n$ in the
power-law modified action of the gravity.

In the following sections, we shall analyze the `dark matter'
problem by considering a generalized version of the virial theorem
in the framework of $f(R)$ modified theories of gravity
\cite{Bohmer:2007fh}. Recall that due to its generality and wide
range of applications, the virial theorem plays an important role
in astrophysics. Assuming steady state, one of the important
results which can be obtained with the use of the virial theorem
is to deduce the mean density of astrophysical objects such as
galaxies, clusters and super clusters, by observing the velocities
of test particles rotating around them. Hence the virial theorem
can be used to predict the total mass of the clusters of galaxies.

The generalized virial theorem, in the context of $f(R)$ gravity
is obtained by using a method based on the collisionless Boltzmann
equation \cite{Bohmer:2007fh}. The additional geometric terms
present in the modified gravitational field equations provide an
effective contribution to the gravitational energy, which at the
galactic/extra-galactic level acts as an effective mass, playing
the role of the `dark matter'. The total virial mass of the
galactic clusters is mainly determined by the effective mass
associated to the new geometrical term, the geometrical mass. It
is important to note that the latter term may account for the
well-known virial theorem mass discrepancy in clusters of
galaxies.

\subsection{The generalized virial theorem in $f(R)$ gravity}
\label{sec:IIIa}

Consider an isolated and spherically symmetric cluster described
by a static and spherically symmetric metric
\begin{equation}
ds^{2}=-e^{\nu \left( r\right) }dt^{2}+e^{\lambda \left( r\right)
}dr^{2}+r^{2}\left( d\theta ^{2}+\sin^{2}\negmedspace\theta
d\varphi ^{2}\right) .  \label{line}
\end{equation}
The galaxies, treated as identical and collisionless point
particles, are described by a distribution function $f_B$, which
obeys the general relativistic Boltzmann equation.

In terms of the distribution function the stress-energy tensor can
be written as $T_{\mu \nu }=\int f_B\, m\, u_{\mu}u_{\nu }\;du$,
where $m$ is the mass of the particle (galaxy)~\cite{Li66},
$u_{\mu }$ is the four-velocity of the galaxy and $du
=du_{r}du_{\theta }du_{\varphi }/u_{t}$ is the invariant volume
element of the velocity space. Thus, the stress-energy tensor of
the matter in a cluster of galaxies can be represented in terms of
an effective density $\rho_{\mathrm{eff}}$ and of an effective
anisotropic pressure, with radial $p_{\mathrm{eff}}^{(r)}$ and
tangential $p_{\mathrm{eff}}^{(\perp)}$ components, given by
\begin{equation}
\rho_{\mathrm{eff}}= \rho \left\langle
u_{t}^{2}\right\rangle,\qquad p_{\mathrm{eff}}^{(r)}=\rho
\left\langle u_{r}^{2}\right\rangle, \qquad
p_{\mathrm{eff}}^{(\perp)}= \rho \left\langle u_{\theta
}^{2}\right\rangle= \rho \left\langle u_{\varphi
}^{2}\right\rangle,
\end{equation}
where, at each point, $\left\langle u_{r}^{2}\right\rangle $ is
the average value of $u_{r}^{2}$, etc, and $\rho $ is the mass
density~\cite{Ja70}.

In what follows, we use this form of the stress-energy tensor, and
for convenience take into account the scalar-tensor representation
of $f(R)$ gravity outlined in Section \ref{sec:scalar-tensor}. As
we are interested in astrophysical applications at the
extra-galactic level, we may assume that the deviations from
standard General Relativity (corresponding to the background value
$\phi =1$) are small. Therefore we may represent $\phi $ as $\phi
=1+\epsilon g^{\prime }(R)$, where $\epsilon $ is a small
quantity, and $g^{\prime }(R)$ describes the modifications of the
geometry due to the presence of the tensor $\theta _{\mu \nu
}$~\cite{Olmo07}, so that $1/\phi \simeq 1-\epsilon g^{\prime
}(R)$. Now adding up the non-zero components of the gravitational
field equation Eq.~(\ref{fieldin}), and taking into account the
above approximations (see Ref. \cite{Bohmer:2007fh} for details),
one obtains the following relationship
\begin{equation}
e^{-\lambda }\left( \frac{\nu ^{\prime \prime }}{2}+\frac{\nu
^{\prime 2}}{4} +\frac{\nu ^{\prime }}{r}-\frac{\nu ^{\prime
}\lambda ^{\prime }}{4}\right) \simeq 4\pi G\rho \left\langle
u^{2}\right\rangle +4\pi G\rho _{\phi }, \label{ff1}
\end{equation}
where $\langle u^{2}\rangle =\langle u_{t}^{2}\rangle +\langle
u_{r}^{2}\rangle +\langle u_{\theta }^{2}\rangle +\langle
u_{\varphi }^{2}\rangle$, and the useful quantity $\rho _{\phi }$
is defined as
\begin{equation}\label{approx}
\rho_{\phi} \simeq -\epsilon \rho \left\langle
u^{2}\right\rangle g^{\prime }(R) \\
+\frac{1}{4\pi G}\left[\frac{1}{\phi }V\left( \phi \right)
+\left.\frac{1}{\phi }\left( 2\nabla _{t}\nabla ^{t}+\square
\right) \phi \right]\right| _{\phi =1+\epsilon g^{\prime }(R)}\,,
\end{equation}
which may be interpreted as the geometric energy density.

It is convenient to introduce some approximations that apply to
test particles in stable circular motion around galaxies, and to
the galactic clusters. First of all, we assume that $\nu $ and
$\lambda $ are slowly varying (i.e.~$\nu^{\prime}$ and
$\lambda^{\prime}$ small), so that in Eq.~(\ref{ff1}) the
quadratic terms can be neglected. Secondly, we assume that the
galaxies have non-relativistic velocities, so that $\langle
u_{1}^{2}\rangle \approx \langle u_{2}^{2}\rangle \approx \langle
u_{3}^{2}\rangle \ll \langle u_{0}^{2}\rangle \approx 1$. Thus,
Eq.~(\ref{ff1}) becomes
\begin{equation}  \label{fin1}
\frac{1}{2r^{2}}\frac{\partial }{\partial r}\left(r^{2}
\frac{\partial \nu}{\partial r}\right) = 4\pi G\rho + 4\pi
G\rho_{\phi}\,.
\end{equation}

In order to derive the virial theorem for galaxy clusters, one
uses the relativistic Boltzmann equation, which provides the
following relationship (see Ref. \cite{Bohmer:2007fh} for details)
\begin{equation}
2K-\frac{1}{2}\int_{0}^{R}4\pi r^{3}\rho \frac{\partial \nu
}{\partial r}dr=0, \qquad {\rm with} \qquad K=\int_{0}^{R}2\pi
\rho \left[ \left\langle u_{1}^{2}\right\rangle +\left\langle
u_{2}^{2}\right\rangle +\left\langle u_{3}^{2}\right\rangle
\right] r^{2}dr\,.
 \label{cond1}
\end{equation}
$K$ is the total kinetic energy of the galaxies, and the total
mass of the system is given by $M=\int_{0}^{R}dM(r)=\int_{0}^{R}
4\pi \rho r^{2}dr$. The main contribution to $M$ is due to the
baryonic mass of the intra-cluster gas and of the stars, but other
particles, such as massive neutrinos, may also contribute
significantly to $M$.

Now, multiplying Eq.~(\ref{fin1}) by $r^{2}$ and integrating from
$0$ to $r$ we obtain
\begin{equation}
GM(r)=\frac{1}{2}r^{2}\frac{\partial \nu }{\partial r}-GM_{\phi
}\left( r\right),  \qquad {\rm with} \qquad  M_{\phi }\left(
r\right) =4\pi \int_{0}^{r}\rho _{\phi}(r')r'^{2} dr'.
 \label{fin2}
\end{equation}
The useful quantity $M_{\phi}$ is denoted as the \textit{geometric
mass} of the cluster. By multiplying Eq.~(\ref{fin2}) with
$dM(r)$, followed by an integration one deduces the relationship
\begin{equation}
\Omega =\Omega _{\phi }-\frac{1}{2}\int_{0}^{R}4\pi r^{3}\rho
\frac{\partial \nu }{\partial r}\,dr\,,
\end{equation}
with the following definitions
\begin{equation}
\Omega =-\int_{0}^{R}\frac{GM(r)}{r}\,dM(r), \qquad {\rm and}
\qquad \Omega _{\phi }=\int_{0}^{R}\frac{GM_{\phi
}(r)}{r}\,dM(r)\,,
\end{equation}
where the quantity $\Omega $ is the usual gravitational potential
energy of the system.

Finally, with the use of Eq.~(\ref{cond1}), we obtain the
generalization of the virial theorem, in $f(R)$ modified theories
of gravity, which takes the form
\begin{equation}
2K + \Omega - \Omega_{\phi } = 0.  \label{theor}
\end{equation}

Note that the generalized virial theorem, given by Eq.
(\ref{theor}), can be written in an alternative form if we
introduce the radii $R_{V}$ and $R_{\phi}$ defined by
\begin{equation}
R_{V}=M^{2}\Bigl/\int_{0}^{R}\frac{M(r)}{r}\,dM(r),\Bigr. \qquad
{\rm and} \qquad R_{\phi }=M_{\phi
}^{2}\Bigl/\int_{0}^{R}\frac{M_{\phi }(r)}{r}\,dM(r)\,,\Bigr.
\label{RU3}
\end{equation}
respectively. We denote $R_{\phi }$ as the \textit{geometric
radius} of the cluster of galaxies. Thus, the quantities $\Omega$
and $\Omega _{\phi}$ are finally given by $\Omega=-GM^2/R_V$ and
$\Omega_\phi=GM_{\phi}^2/R_{\phi}$, respectively.

The virial mass $M_{V}$ is defined as
\begin{equation}
2K=\frac{GMM_{V}}{R_{V}}\,.
\end{equation}
After substitution into the virial theorem, given by
Eq.~(\ref{theor}), we obtain
\begin{equation}  \label{fin6}
\frac{M_{V}}{M}=1+\frac{M_{\phi }^{2}R_{V}}{M^{2}R_{\phi }}\,.
\end{equation}
If $M_{V}/M>3$, a condition which is valid for most of the
observed galactic clusters, then Eq.~(\ref{fin6}) provides the
virial mass in $f(R)$ gravity, which can be approximated by
\begin{equation}
M_{V}\approx \frac{M_{\phi}^2}{M}\frac{R_{V}}{R_{\phi }}\,.
\label{virial}
\end{equation}

\subsection{Geometric mass and geometric radius from galactic
cluster observations}\label{sec:IIIb}

An interesting application of the generalized virial theorem can
be inferred from the galaxy cluster observations. According to the
modified $f(R)$ gravity model, the total mass of the cluster
consists of the sum of the baryonic mass (mainly the intra-cluster
gas), and the geometric mass, so that $M_{tot}(r)=4\pi
\int_{0}^{r}\left( \rho _{g}+\rho _{\phi }\right) r^{2}dr$. Hence
it follows that $M_{tot}(r)$ satisfies the following mass
continuity equation
\begin{equation}
\frac{dM_{tot}\left( r\right) }{dr}=4\pi r^{2}\rho _{g}\left(
r\right) +4\pi r^{2}\rho _{\phi }\left( r\right) .  \label{massf}
\end{equation}

Note that most of the baryonic mass in the clusters of galaxies is
in the form of the intra-cluster gas. The gas mass density $\rho
_{g}$ distribution can be fitted with the observational data by
using the following expression for the radial baryonic mass (gas)
distribution~\cite{ReBo02}
\begin{equation}
\rho _{g}(r)=\rho _{0}\left( 1+\frac{r^{2}}{r_{c}^{2}}\right)
^{-3\beta /2}, \label{dens}
\end{equation}
where $r_{c}$ is the core radius, and $\rho _{0}$ and $\beta $ are
(cluster-dependent) constants. Using the Jeans
equation~\cite{Bi87}, one may obtain the total mass
distribution~\cite{ReBo02,HaCh07}, so that taking into account the
density profile of the gas given by Eq.~(\ref {dens}), the total
mass profile inside the cluster is given by
\begin{equation}
M_{tot}(r)=\frac{3k_{B}\beta T_{g}}{\mu
m_{p}G}\frac{r^{3}}{r_{c}^{2}+r^{2}} \label{mp} \,,
\end{equation}
where $k_{B}$ is Boltzmann's constant, $T_{g}$ is the gas
temperature, $\mu \approx 0.61$ is the mean atomic weight of the
particles in the cluster gas, and $m_{p}$ is the proton
mass~\cite{ReBo02}.

Using the mass continuity equation, Eq. (\ref{massf}), then
Eqs.~(\ref{dens}) and (\ref{mp}) provide the expression of the
geometric density term inside the cluster, given by
\begin{equation}
4\pi \rho _{\phi }\left( r\right)=\frac{3k_{B}\beta T_{g}\left(
r^{2}+3r_{c}^{2}\right) }{\mu m_{p}\left( r_{c}^{2}+r^{2}\right)
^{2}}-\frac{4\pi G\rho _{0}}{\left( 1+r^{2}/r_{c}^{2}\right)
^{3\beta /2}}\,.
\end{equation}
In the limit $r\gg r_{c}$ we obtain for $\rho _{\phi }$ the simple
relation
\begin{equation}
4\pi \rho _{\phi }\left( r\right)=\left[ \frac{3k_{B}\beta
T_{g}}{\mu m_{p}}-4\pi G\rho _{0}r_{c}^{3\beta }r^{2-3\beta
}\right] \frac{1}{r^{2}}\,,
\end{equation}
and the geometric mass in the limit $r\gg r_{c}$, may be
approximated as
\begin{equation}
GM_{\phi }\left( r\right) \approx \left[ \frac{3k_{B}\beta
T_{g}}{\mu m_{p}}-\frac{4\pi G\rho _{0}r_{c}^{3\beta }r^{2-3\beta
}}{3\left( 1-\beta \right) } \right] r\,.  \label{GM}
\end{equation}

One may assume that the contribution of the gas density and mass
to the geometric density and geometric mass, respectively, can be
neglected. The latter approximations are very well supported by
astrophysical observations, which show that the gas represents
only a small fraction of the total mass \cite{ReBo02,Ar05}.
Therefore, we obtain
\begin{equation}\label{rhophi}
4\pi G\rho _{\phi }(r)\approx \left( \frac{3k_{B}\beta T_{g}}{\mu
m_{p}} \right) r^{-2}, \qquad {\rm and} \qquad GM_{\phi }\left(
r\right) \approx \left( \frac{3k_{B}\beta T_{g}}{\mu m_{p}}
\right) r,
\end{equation}
respectively.

One may also estimate an upper bound for the cutoff of the
geometric mass. The idea is to consider the point at which the
decaying density profile of the geometric density associated to
the galaxy cluster becomes smaller than the average energy density
of the Universe. Let the value of the coordinate radius at the
point where the two densities are equal to be $R_{\phi }^{(cr)}$.
Then at this point $\rho _{\phi }(R_{\phi}^{(cr)})=\rho_{univ}$,
where $\rho _{univ}$ is the mean energy density of the universe.
By assuming $\rho_{univ}=\rho _{c}=3H^{2}/8\pi
G=4.6975\times10^{-30}h_{50}^{2}\;\mathrm{g}/\mathrm{cm}^{-3}$,
where $H=50h_{50}\;\mathrm{km}/\mathrm{Mpc}/\mathrm{s}$~
\cite{ReBo02}, we obtain
\begin{eqnarray}
R_{\phi }^{(cr)}=\left( \frac{3k_{B}\beta T_{g}}{\mu m_{p}G\rho
_{c}} \right)^{1/2}  =91. 33\sqrt{\beta }\left(
\frac{k_{B}T_{g}}{5\text{keV}} \right)
^{1/2}h_{50}^{-1}\mathrm{Mpc.}  \label{Rucr}
\end{eqnarray}

The total geometric mass corresponding to this value is
\begin{eqnarray}
M_{\phi }^{(cr)}=M_{\phi }\left( R_{\phi }^{(cr)}\right)
=4.83\times 10^{16}\beta ^{3/2}\left(
\frac{k_{B}T_{g}}{5\text{keV}} \right) ^{3/2}h_{50}^{-1}M_{\odot
}.
\end{eqnarray}
This value of the mass is consistent with the observations of the
mass distribution in the clusters of galaxies. However, according
to $f(R)$ modified theories of gravity, we predict that the
geometric mass and its effects extends beyond the virial radius of
the clusters, which is of the order of only a few Mpc.

By assuming that $R_{\phi }\approx R_{\phi }^{(cr)}$, we obtain
the following relation between the virial and the baryonic mass of
the cluster
\begin{equation}
M_{V}\approx 91.33\sqrt{\beta }\left( \frac{k_{B}T_{g}}{5\;\mathrm{keV}}%
\right) ^{1/2}h_{50}^{-1}\frac{M}{R_{V}(\mathrm{Mpc})}.
\end{equation}
For a cluster with gas temperature $T_{g}=5\times 10^{7}$ K,
$\beta =1/2$ and $R_{V}=2$ Mpc we obtain $M_{V}\approx 32M$, a
relation which is consistent with the astronomical observations
\cite{ReBo02}.

\section{Summary and discussion}\label{sec:conclusion}\label{sec:IV}

Cosmology has entered a `golden age', in which the rapid
development of increasingly high-precision data has turned it from
a speculative to an observationally based science. Recent
experiments call upon state of the art technology to provide
detailed information about the contents and history of the
Universe. These experiments include the Hubble Space Telescope,
the NASA WMAP satellite instrument, that measures the temperature
and polarisation of the CMBR, and the Sloan Digital Sky Survey
(SDSS), that is automatically mapping the properties and
distribution of 1 million galaxies. High-precision cosmology has
allowed us to tie down the parameters that describe our Universe
with growing accuracy.

The standard model of cosmology is remarkably successful in
accounting for the observed features of the Universe. However,
there remain a number of fundamental open questions at the
foundation of the standard model. In particular, we lack a
fundamental understanding of the acceleration of the late
universe. Recent observations of supernovae, together with the
WMAP and SDSS data, lead to the remarkable conclusion that our
universe is not just expanding, but has begun to
accelerate~\cite{expansion}. What is the so-called `dark energy'
that is driving the acceleration of the universe? Is it a vacuum
energy or a dynamical field (``quintessence'')? Or is the
acceleration due to infra-red modifications of Einstein's theory
of General Relativity? How is structure formation affected in
these alternative scenarios? What will the outcome be of this
acceleration for the future fate of the universe?

The aspects of these fundamental questions whose resolution is so
important for theoretical cosmology, need to look beyond the
standard theory of gravity. It is clear that these questions
involve not only gravity, but also particle physics. String theory
provides a synthesis of these two parts of physics and is widely
believed to be moving towards a viable quantum gravity theory. One
of the key predictions of string theory is the existence of extra
spatial dimensions. In the brane-world scenario, motivated by
recent developments in string theory, the observed 3-dimensional
universe is embedded in a higher-dimensional
spacetime~\cite{Maartens1}. The generalization of the
Dvali-Gabadadze-Porrati (DGP) brane models \cite{DGP} lead to
late-accelerating cosmologies \cite{Deffayet}, even in the absence
of a dark energy field.

This exciting feature of ``self acceleration'' may help towards a
new resolution to the dark energy problem, although this model
deserves further investigation as a viable cosmological model
\cite{Maartens2}. It will be interesting to generalize the DGP
model with the inclusion of a Gauss-Bonnet (GB) \cite{Maartens3},
and it will also be important to investigate the effects of the GB
term relatively to the issues of strong coupling and ghosts in the
DGP models. Infra-red modifications to General Relativity, where
the consistency of various candidate models, including
4-dimensional modifications to the Einstein-Hilbert action,
especially GB modifications with a scalar field coupling, have
also be analyzed \cite{Nojiri:2005vv}. In this context, a more
general modification of the Einstein-Hilbert gravitational
Lagrangian density in the form of $L=f(R)$ has recently been
extensively analyzed.

Relatively to the construction of ``quintessential'' dark energy
models, recent fits to observational data indicate that an
evolving equation of state crossing the phantom divide is mildly
favored \cite{Zhang,Upadhye}. In a cosmological setting, it has
also been shown that the transition into the phantom regime, a
mixture of various interacting non-ideal fluids is necessary
\cite{Vikman}, with important implications to the model
construction of dark energy. If confirmed in the future, this
behaviour holds important implications to the model construction
of dark energy. The latter models, considering a redshift
dependent equation of state, possibly provide better fits to the
most recent and reliable SN Ia supernovae Gold dataset.

Deciding between these possible sources of the cosmic acceleration
will be one of the major objectives in cosmology in the next
decade with several surveys and experiments to address the nature
of dark energy. One may mention new several major SNIa supernovae
projects, such as the SuperNova Legacy Survey (SNLS), SDSS-II,
Destiny, the Large Synoptic Survey Telescope (LSST), Dark Energy
Survey (DES) and the SuperNova Acceleration Probe (SNAP). Other
dark energy probes include the Dark UNiverse Explorer (DUNE); the
Wide Field Multi-Object Spectrograph (WFMOS), which will perform
surveys to measure dark energy and the history of our Galaxy; and
the Panoramic Survey Telescope and Rapid Response System
(PanSTARRS), amongst others. All of these aspects present an
extremely fascinating aspect for the above-mentioned experiments
and for future theoretical research.

\acknowledgments

I thank Orfeu Bertolami, Christian Boehmer, Tiberiu Harko, Kazuya
Koyama, Roy Maartens and Antonios Papazoglou for helpful comments,
and acknowledge funding by Funda\c{c}\~{a}o para a Ci\^{e}ncia e a
Tecnologia (FCT)--Portugal through the grant SFRH/BPD/26269/2006.

\newpage


\end{document}